\documentclass[a4paper,11pt]{article}
\usepackage{pos}
\usepackage{mathrsfs}
\usepackage{braket}
\usepackage{bm}
\usepackage{comment}

\title{Restoration of residual gauge symmetries due to topological defects and color confinement in the Lorenz gauge}
\ShortTitle{Restoration of RGS due to topological defects and color confinement}

\author*[a]{Naoki Fukushima}
\author[b]{Kei-Ichi Kondo}

\affiliation[a]{Department of Physics, Graduate School of Science and Engineering,\\ Chiba University, Chiba 263-8522, Japan}

\affiliation[b]{Department of Physics, Graduate School of Science,\\ Chiba University, Chiba 263-8522, Japan}

\emailAdd{fnuakouksih9i1m3a@chiba-u.jp}
\emailAdd{kondok@faculty.chiba-u.jp}

\abstract{
The residual gauge symmetry (RGS) is the local gauge symmetry remaining even after imposing the gauge fixing condition. Although this symmetry is “spontaneously broken” in the perturbative vacuum, it can be restored in the true confining vacuum of QCD. Therefore, a color confinement criterion is obtained as the condition of restoration of the RGS, namely, disappearance of the massless Nambu-Goldstone pole associated with this spontaneous breaking, provided that the color confinement phase is a disordered phase where all internal symmetries remain unbroken. In the Lorenz gauge, indeed, it was shown by Hata that the restoration condition is identical to the Kugo-Ojima color confinement criterion, if the gauge transformation function $\omega (x)$ for the residual gauge symmetry is taken to be linear in $x$. However, this result was obtained without regard to topological configurations.
In this talk, we reconsider this issue by taking into account topological defects that are expected to play the dominant role for realizing confinement in the non-perturbative way.
}

\FullConference{The XVIth Quark Confinement and the Hadron Spectrum Conference (QCHSC24)\\
 19-24 August, 2024\\
 Cairns Convention Centre, Cairns, Queensland, Australia\\}

\begin{document}
\maketitle

\section{Introduction}
     Quark confinement is well understood within the framework of the dual superconductor picture, where the condensation of magnetic monopoles and antimonopoles occurs \cite{dual-superconductor}, which is also developed from a gauge independent perspective \cite{dual-superconductor-Kondo}.
     However, in more general cases including gluons and composite particles such as hadrons and glueballs, the dual superconductor picture provides less clear understanding.

     In this context, we revisit the Kugo-Ojima (KO) color confinement criterion proposed by Kugo and Ojima \cite{KO}: if the KO criterion is satisfied, all colored objects cannot be observed. However, the KO criterion was derived only in the Lorenz gauge $\partial^\mu \mathscr{A}_\mu = 0$ and cannot be applied to the other gauge-fixing conditions.

     In the Lorenz gauge, the KO criterion is also derived as the condition of the restoration of the residual gauge symmetry (RGS) with the special choice of the gauge transformation function $\omega (x)$ by Hata \cite{Hata-method}. The RGS is the local gauge symmetry remaining even after imposing the gauge fixing condition.
     This symmetry is ``spontaneously broken'' in the perturbative vacuum.
     Since we consider the color confinement phase to be a disordered phase where all of symmetries are unbroken, the RGS should be also restored in the true confining vacuum of QCD. Then, we can consider the condition of the restoration as the disappearance of the massless Nambu-Goldstone(NG) pole associated with this sponteneously breaking.

     We try to generalize the condition of the restoration of the RGS into more general case including topological configurations \cite{Kondo-Fukushima}.
     To correctly take into account topological configurations, we must consider a finite gauge transformation, although we have used the infinitesimal gauge transformation in the previous work. In this talk, we reconsider the condition of the restoration of the RGS by taking into account topological configurations.

\section{The restoration condition of the RGS in the Lorenz gauge}
     In the manifestly Lorentz covariant operator formalism on the indefinite metric state space $\mathcal{V}$, we suppose that the (nilpotent) BRST symmetry exists.
     Let $\mathcal{V}_{\text{phys}}$ be the physical state space with a semi-positive definite metric $\braket{\text{phys} | \text{phys}} \geq 0$, which is defined by using the BRST charge $Q_{\text{B}}$:
\begin{align}
  \mathcal{V}_{\text{phys}} = \{\ket{\text{phys}} \in \mathcal{V} ; Q_{\text{B}} \ket{\text{phys}} = 0\} \subset \mathcal{V}.
  \label{eq:physical_state}
\end{align}
     Here, we choose the Lorenz gauge fixing:
\begin{align}
  \partial^\mu \mathscr{A}_\mu^A (x) = 0.
\end{align}
     Then, the Lagrangian density in the Lorenz gauge is given by using BRST transformation $\delta_{\text{B}}$ as
\begin{align}
  \mathscr{L} &= \mathscr{L}_{\text{YM}} + \mathscr{L}_{\text{GF+FP}}, \nonumber\\
  \mathscr{L}_{\text{YM}} &= - \frac{1}{4} \mathscr{F}_{\mu \nu}^A \mathscr{F}^{\mu \nu A} + \frac{i \vartheta}{32 \pi^2} \mathscr{F}_{\mu \nu}^A {}^* \mathscr{F}^{\mu \nu A} , \ \mathscr{L}_{\text{GF+FP}} = - i \delta_{\text{B}} \left\{\bar{\mathscr{C}}^A \left(\partial^\mu \mathscr{A}_\mu^A + \frac{\alpha}{2} \mathscr{B}^A\right)\right\},
\end{align}
     where $\vartheta$ is the topological angle. We consider the generalized (finite) local gauge transformation $U (x)$:
\begin{align}
  \delta_U \mathscr{A}_\mu (x) =& \Omega_\mu (x) + U (x) \mathscr{A}_\mu (x) U^\dagger (x) - \mathscr{A}_\mu (x) , \ \Omega_\mu (x) := i g^{- 1} U (x) \partial_\mu U^\dagger (x) ,\nonumber\\
  \delta_U \mathscr{B} (x) = &U (x) \mathscr{B} (x) U^\dagger (x) - \mathscr{B} (x), \nonumber\\
  \delta_U \mathscr{C} (x) =& U (x) \mathscr{C} (x) U^\dagger (x) - \mathscr{C} (x), \ \delta_U \bar{\mathscr{C}} (x) = U (x) \bar{\mathscr{C}} (x) U^\dagger (x) - \bar{\mathscr{C}} (x).
  \label{eq:gauge_transformation}
\end{align}
    For $U (x)$ to be a RGS for the Lorenz gauge, $U (x)$ should satisfy
\begin{align}
  0 = \partial^\mu \delta_U \mathscr{A}_\mu (x) \ \Leftrightarrow \ 0 = \partial^\mu (\Omega_\mu (x) + U (x) \mathscr{A}_\mu (x) U^\dagger (x)).
  \label{eq:Laplace}
\end{align}
The infinitesimal version $U = e^{i g \omega (x)} \approx 1 + i g \omega (x)$ is given by
\begin{align}
  \delta_\omega \mathscr{A}_\mu (x) = \mathscr{D}_\mu \omega (x) , \ \delta_\omega \mathscr{B} (x) = g (\mathscr{B} \times \omega), \ \delta_\omega \mathscr{C} (x) = g (\mathscr{C} \times \omega), \ \delta_\omega \bar{\mathscr{C}} (x) = g (\bar{\mathscr{C}} \times \omega).
  \label{eq:infinitesimal_gt}
\end{align}

    In discussing the Nambu-Goldstone pole (and the restoration of the RGS), it is enough to consider the infinitesimal gauge transformation.
     We can calculate the Noether current $\mathscr{J}_\omega^\mu$ associated with \eqref{eq:infinitesimal_gt}. Then the divergence of this current is equal to the transformation of the Lagrangian density under the transformation \eqref{eq:infinitesimal_gt}
\begin{align}
  \partial_\mu \mathscr{J}_\omega^\mu &= \delta_\omega \mathscr{L} = \delta_\omega \mathscr{L}_{\text{GF+FP}}
  = - i \delta_\omega \delta_{\text{B}} \left[\bar{\mathscr{C}}^A \left(\partial^\mu \mathscr{A}_\mu^A + \frac{\alpha}{2} \mathscr{B}^A\right)\right] \nonumber\\
  &= - i \delta_{\text{B}} \delta_\omega \left[\bar{\mathscr{C}}^A \left(\partial^\mu \mathscr{A}_\mu^A + \frac{\alpha}{2} \mathscr{B}^A\right)\right] = - i \delta_{\text{B}} \delta_\omega [\bar{\mathscr{C}}^A \partial^\mu \mathscr{A}_\mu^A] = i \delta_{\text{B}} [(\mathscr{D}_\mu \bar{\mathscr{C}})^A] \partial^\mu \omega^A ,
  \label{eq:L_gauge_transform_Lorenz}
\end{align}
     where we have used the fact that the gauge transformtion and the BRST transformation commute \cite{Kondo-Fukushima}. 
     Therefore, the local current $\mathscr{J}_\omega^\mu$ \textit{is conserved in the physical state space} $\mathcal{V}_{\text{phys}}$ according to \eqref{eq:physical_state}, because \eqref{eq:L_gauge_transform_Lorenz} is BRST exact. We follow the idea of Hata for enlarging the concept of the symmetry by combining the state in the quantum theory without being restricted to the Noether method in the classical theory.

     The restoration condition of the RGS is written as the condition of the disappearance of the massless Nambu-Goldstone pole. In a single gauge field sector $\Phi = \mathscr{A}_\nu^B (y)$, we focus on the following Ward-Takahashi (WT) identity
\begin{align}
  i \partial_\mu^x \braket{\mathscr{J}_\omega^\mu (x) \Phi}
  = \delta^D (x - y) \Braket{\delta_\omega \mathscr{A}_\nu^B (x)} + i \braket{\partial_\mu \mathscr{J}_\omega^\mu (x) \mathscr{A}_\nu^B (y)}.
\end{align}
     Performing the Fourier transformation, we obtain
\begin{align}
  i \int d^D x \ e^{i p (x - y)} \partial_\mu^x \braket{\mathscr{J}_\omega^\mu (x) \mathscr{A}_\nu^B (y)} = \braket{\delta_\omega \mathscr{A}_\nu^B (y)} + i \int d^D x \ e^{i p (x - y)} \braket{\delta_\omega \mathscr{L} (x) \mathscr{A}_\nu^B (y)}.
  \label{eq:A-WI-1_Lorenz}
\end{align}
     From \eqref{eq:L_gauge_transform_Lorenz}, the last term of \eqref{eq:A-WI-1_Lorenz} is reduced to
\begin{align}
  &i \int d^D x \ e^{i p (x - y)} \braket{\delta_\omega \mathscr{L} (x) \mathscr{A}_\nu^B (y)}
  = - \int d^D x \ e^{i p (x - y)} \partial^\mu \omega^A (x) \braket{(\mathscr{D}_\mu \bar{\mathscr{C}})^A (x) \delta_{\text{B}} \mathscr{A}_\nu^B (y)} \nonumber\\
  = &- \int d^D x \ e^{i p (x - y)} \partial^\mu \omega^A (x) \left[\frac{\partial^x_\mu \partial^x_\nu}{\partial_x^2} \delta^D (x - y) + \left(g_{\mu \rho} - \frac{\partial^x_\mu \partial^x_\rho}{\partial_x^2}\right) \braket{g (\mathscr{A}^\rho \times \bar{\mathscr{C}})^A (x) (\mathscr{D}_\nu \mathscr{C})^B (y)}\right],
  \label{eq:WI-second_term_Lorenz}
\end{align}
    where we have used $\braket{\delta_{\text{B}} [\cdots]} = 0$ and the Schwinger-Dyson equation $\int d \mu \ \frac{\delta}{\delta \Phi (x)} (e^{i S} F) = 0$.

     Thus we obtain the restoration condition of the RGS in the single gauge field sector $\mathscr{A}_\nu^B (y)$ in the $p \rightarrow \ 0$ limit of \eqref{eq:A-WI-1_Lorenz}: \cite{Fukushima-Kondo}
\begin{align}
  I_\nu^B
  &:= \lim_{p \rightarrow \  0} \int d^D x \ e^{i p (x - y)} \partial^\mu \omega^A (x) \left(g_{\mu \nu} - \frac{\partial^x_\mu \partial^x_\nu}{\partial_x^2} \right) (\delta^{A B} \delta^D (x - y) + u^{A B} (x - y)) \nonumber\\
  &
  \begin{cases}
    = 0 &\text{restoration} \\
    \neq 0 &\text{no restoration}
  \end{cases}
  .
  \label{eq:condition-A-Lorenz}
 \end{align}
 where we have defined the Kugo-Ojima (KO) function $u^{A B}$ in the configuration space
\begin{align}
  \left(g_{\mu \rho} - \frac{\partial^x_\mu \partial^x_\rho}{\partial_x^2}\right) \braket{g (\mathscr{A}^\rho \times \bar{\mathscr{C}})^A (x) (\mathscr{D}_\nu \mathscr{C})^B (y)} := - \left(g_{\mu \nu} - \frac{\partial^x_\mu \partial^x_\nu}{\partial_x^2}\right) u^{A B} (x - y).
\end{align}
     In the non-compact gauge theory, it is enough to consider the infinitesimal gauge transformation and the RGS $\partial^\mu \omega^A (x) = \delta^{A C} b^\mu$ ($C$ is an arbitrary component of the Lie algebra).
     Indeed, \eqref{eq:condition-A-Lorenz} is reduced to the Kugo-Ojima criterion as shown by Hata \cite{Hata-method}
\begin{align}
  0 = \lim_{p^2 \rightarrow \ 0} [\delta^{A B} + u^{A B} (p^2)].
  \label{eq:KO}
\end{align}

\section{Construction of the residual gauge symmetry}
\subsection{Witten Ansatz}
For concreteness, we restrict the gauge group to $SU (2)$ and consider the Euclidean spacetime. The element $U = e^{i \theta n^A \frac{\sigma_A}{2}}$ of $SU (2)$ group is written as
\begin{align}
  U (x) = \exp \left(i \theta (x) \text{n}^A (x) \frac{\sigma_A}{2}\right) = \cos \frac{\theta (x)}{2} + i \sin \frac{\theta (x)}{2} \text{n}^A (x) \sigma_A \ (A = 1 , 2 , 3) ,
\end{align}
where $\mathbf{n} (x)$ is a three-dimensional unit vector: $\mathbf{n} (x) \cdot \mathbf{n} (x) = \text{n}^A (x) \text{n}^A (x) = 1$ and $\theta (x)$ is an angle of the rotation around $\mathbf{n} (x)$. The pure gauge form is given by
\begin{align}
  \Omega_\mu (x) := i U (x) \partial_\mu U^\dagger (x) = \frac{\bm{\sigma}}{2} \cdot [\partial_\mu \theta (x) \mathbf{n} (x) + \sin \theta (x) \partial_\mu \mathbf{n} (x) + (1 - \cos \theta (x)) (\partial_\mu \mathbf{n} (x) \times \mathbf{n} (x))].
\end{align}
If we require the spherical symmetry in space $\mathbb{R}^3$ and adopt
\begin{align}
  \text{n}^A = \frac{x_A}{r} , \ r := \sqrt{x_1^2 + x_2^2 + x_3^2},
\end{align}
we obtain
\begin{align}
  \Omega_0 (x) = i U \partial_0 U^\dagger &= \frac{\sigma_A}{2} \frac{x_A}{r} \frac{\partial \theta}{\partial t} , \nonumber\\
  \Omega_j (x) = i U \partial_j U^\dagger &= \frac{\sigma_A}{2} \left\{  \frac{x_j x_A}{r^2} \frac{\partial \theta}{\partial r}  + \frac{\delta_{j A} r^2 - x_j x_A}{r^3} \sin \theta - \frac{\varepsilon_{j A k} x_k}{r^2} [1 - \cos \theta]  \right\} .
  \label{eq:Omega_Witten_Ansatz}
\end{align}
This obserbation suggests the Witten Ansatz \cite{Witten1977} for the $SU (2)$ Yang-Mills field with cylindrical symmetry in $\mathbb{R}^4$:
\begin{align}
  -\mathscr{A}_4^A (x) &= \frac{x_A}{r} A_0 (r , t) , \nonumber\\
  -\mathscr{A}_j^A (x) &= \frac{x_j x_A}{r^2} A_1 (r , t) + \frac{\delta_{j A} r^2 - x_j x_A}{r^3} \varphi_1 (r , t)  + \frac{\varepsilon_{j A k} x_k}{r^2} [1 + \varphi_2 (r , t)] \ (j = 1 , 2 , 3),
  \label{eq:Witten_Ansatz}
\end{align}
where $t$ is the Euclidean time and $r$ is the spatial radius.
Though the Witten Ansatz was originally introduced to derive the multi-instanton solution of the self-dual equation $\mathscr{F}_{\mu \nu} = \pm {}^* \mathscr{F}_{\mu \nu}$ in the $D = 4$ Euclidean space $(t , x_1 , x_2 , x_3)$, we use this Ansatz to simplify the equation \eqref{eq:Laplace} for discussing existence of the RGS.

Under this Ansatz, the Euclidean Yang-Mills Lagrangian density is rewritten in terms of the Ansatz functions into
\begin{align}
  \mathscr{L}_{\text{YM}} &= \frac{1}{4} \mathscr{F}_{\mu \nu}^A (x) \mathscr{F}_{\mu \nu}^A (x) + \frac{\vartheta}{32 \pi^2} \mathscr{F}_{\mu \nu} {}^* \mathscr{F}_{\mu \nu}
  = \frac{1}{2} (\mathscr{F}_{4 j}^A)^2 + \frac{1}{2} \left(\frac{1}{2} \varepsilon_{jk \ell} \mathscr{F}_{k \ell}^A\right)^2 + \frac{\vartheta}{32 \pi^2} \mathscr{F}_{\mu \nu} {}^* \mathscr{F}_{\mu \nu} \nonumber\\
  &= \frac{1}{r^2} D_\mu \varphi_a D_\mu \varphi_a + \frac{1}{4} F_{\mu \nu} F_{\mu \nu} + \frac{1}{2 r^4}(1 - \varphi_a \varphi_a)^2  + \frac{\vartheta}{16 \pi^2 r^2} \varepsilon_{\mu \nu} F_{\mu \nu},
  \label{YM-L}
\end{align}
where we have ignored the total derivative of the topological term \cite{Witten1977} and used the fact that the field strength $\mathscr{F}_{\mu\nu}^A$ is given by defining $F_{\mu \nu} := \partial_\mu A_\nu - \partial_\nu A_\mu , \ D_\mu \varphi_a := \partial_\mu \varphi_a + \varepsilon_{a b} A_\mu \varphi_b \ (\mu , \nu = 0 , 1 , \ a , b = 1 , 2)$:
\begin{align}
  - \mathscr{F}_{4 j}^A = &  \frac{x_j x_A}{r^2} F_{0 1}  +  \frac{\delta_{j A} r^2 - x_j x_A}{r^3} D_0 \varphi_1 + \frac{\varepsilon_{j A k} x_k}{r^2} D_0 \varphi_2 ,  \nonumber\\
   - \frac{1}{2} \varepsilon_{jk \ell} \mathscr{F}_{k \ell}^A  = &  \frac{x_j x_A }{r^2} \frac{\varphi_1^2 + \varphi_2^2-1}{r^2}  -  \frac{\delta_{j A} r^2 - x_j x_A }{r^3} D_1 \varphi_2 + \frac{\varepsilon_{j A k} x_k}{r^2} D_1 \varphi_1 .
   \label{F}
\end{align}
\eqref{YM-L} indicates that the dimensional reduction from the $D = 4$ $SU (2)$ Yang-Mills theory to the $D = 2$ (with coordinate $(r , t)$) $U(1)$ gauge-scalar theory occurs. In this dimensional reduction, the gauge transformation $U = e^{i \theta \frac{x_A}{r} \frac{\sigma_A}{2}} \in SU (2)$ is replaced by \cite{Actor79}
\begin{align}
  A_\mu \rightarrow \ A_\mu^\prime = A_\mu - \partial_\mu \theta , \
  \begin{pmatrix}
    \varphi_1 \\
    \varphi_2
  \end{pmatrix}
  \rightarrow \
  \begin{pmatrix}
    \varphi_1^\prime \\
    \varphi_2^\prime
  \end{pmatrix}
=  \begin{pmatrix}
    \cos \theta & \sin \theta \\
    - \sin \theta & \cos \theta
  \end{pmatrix}
  \begin{pmatrix}
    \varphi_1 \\
    \varphi_2
  \end{pmatrix}
  .
  \label{eq:gauge_transformation_Witten_Ansatz}
\end{align}
This is nothing but a finite $U(1)$ gauge transformation.

\subsection{Residual gauge symmetry}
The Lorenz gauge condition under the Witten Ansatz is reduced to
\begin{align}
  \partial_\mu (r^2 A_\mu(r , t)) = 2 \varphi_1(r , t).
  \label{original-gfc}
\end{align}
In order for $U = e^{i \theta \frac{x_A}{r} \frac{\sigma_A}{2}}$ to satisfy \eqref{eq:Laplace}, $\theta$ must satisfy
\begin{align}
  &\partial_\mu [r^2 (A_\mu(r , t) - \partial_\mu \theta(r , t))] = 2 (\cos \theta(r , t) \varphi_1(r , t) + \sin \theta(r , t) \varphi_2(r , t)) \nonumber\\
  \Rightarrow \ &\partial_\mu (r^2 \partial_\mu \theta (r , t))
    =  2(1-\cos \theta (r , t)) \varphi_1 (r , t) -2 \sin \theta (r , t) \varphi_2 (r , t) \ (\mu = 0 , 1)  .
  \label{gfc}
\end{align}
This is nothing but the Gribov equation \cite{Gribov}.
For discussing spontaneous symmetry breaking, it is enough to consider the vacuum configuration $\mathscr{A}_\mu \equiv 0$.
This corresponds to $A_0 = A_1 = \varphi_1 = 0$ and $\varphi_2 = - 1$. Therefore, \eqref{gfc} reads
\begin{align}
  \partial_\mu (r^2 \partial_\mu \theta (r , t)) - 2 \sin \theta (r , t) = 0.
  \label{eq:gfc-perturbative0}
\end{align}
When $\theta$ is independent of $t$, this is simplified to
\begin{align}
  \partial_1 (r^2 \partial_1 \theta (r)) - 2 \sin \theta (r) = 0.
  \label{eq:gfc-perturbative}
\end{align}
If we define $\tau := \ln r$ and denote the differentiation with respect to $\tau$ by dot, we can rewrite \eqref{eq:gfc-perturbative}
\begin{align}
  \ddot{\theta} (\tau) + \dot{\theta} (\tau) - \sin \theta (\tau) = 0
\end{align}
This is nothing but the equation of motion of the Gribov pendulum and the solution for $\theta (r)$ indeed exists as shown by Gribov \cite{Gribov}.

In order to linearize the equation \eqref{eq:gfc-perturbative} around $\theta = \theta_0$, we substitute $\theta (r) = \theta_0 + \delta (r)$ and extract the linear term in $\delta$:
\begin{align}
  r^2 \delta^{\prime \prime}(r) + 2 r \delta^{\prime}(r) - 2 \cos \theta_0 \delta(r) = 2 \sin \theta_0,
\end{align}
where the prime denotes the differentiation with respect to $r$. This is the second order linear differential equation of the Euler type. Therefore, the asymptotic solution for $r \rightarrow \ 0$ or $r \rightarrow \ \infty$ is obtained as follows:
\begin{align}
  \theta (r) \sim
  \begin{cases}
    \theta_0 - \tan \theta_0 + C_1 r^{\frac{1}{2}(- 1 + \sqrt{1 + 8 \cos \theta_0})} &(1 + 8 \cos \theta_0 > 0 , \ r \rightarrow \ 0) \\
    \theta_0 - \tan \theta_0 + C_2 r^{- \frac{1}{2}} \cos \left(\frac{1}{2}\sqrt{|1 + 8 \cos \theta_0|} \ln r + \alpha\right) & (1 + 8 \cos \theta_0 < 0 , \ r \rightarrow \ \infty)
  \end{cases}
  .
\end{align}
Since \eqref{eq:gfc-perturbative} is a second-order differential equation, its solution $\theta (r)$ contains infinitesimal global parameters, such as $C_1$ and $C_2$, which play the role of defininig the global symmetry according to the Noether method, corresponding to $b^\mu$ in Hata's example: $\partial^\mu \omega^A (x) = \delta^{A C} b^\mu$.

\subsection{Finite action and contribution to the path integral}
Under the Witten Ansatz, $D = 4$ Euclidean Yang-Mills action $S_{\text{YM}} = \int d^4 x \ \mathscr{L}_{\text{YM}}$ is rewritten by
\begin{align}
  S_{\text{YM}} = 4 \pi \int_{- \infty}^\infty d t \int_0^\infty d r \ \left[\frac{1}{2} D_\mu \varphi_a D_\mu \varphi_a + \frac{1}{8} r^2 F_{\mu \nu} F_{\mu \nu} + (1 - \varphi_a \varphi_a)^2 \frac{1}{4 r^2} + \frac{\vartheta}{16 \pi^2} \varepsilon_{\mu \nu} F_{\mu \nu}\right] .
  \label{eq:YM2_action}
\end{align}
The Ansatz functions $A_0$, $A_1$, $\varphi_1$ and $\varphi_2$ must be chosen to give a finite Euclidean action $S_{\text{YM}}<\infty$ so as to contribute to the path integral.

When the topological configuration is independent of $t$, in order to obtain the finite Euclidean action (without integration over $t$ variable), we must restrict the Ansatz functions to the function satisfying the following boundary conditions at $r = 0$ and $r = \infty$:
\begin{align}
  &
  \begin{cases}
    D_\mu \varphi_a = \mathcal{O} \left(\left(\frac{1}{r}\right)^{\frac{1}{2} + \delta_1}\right) , \ F_{\mu \nu} = \mathcal{O} \left(\left(\frac{1}{r}\right)^{\frac{3}{2} + \delta_2}\right) & (r \rightarrow \ \infty , \ 0 < \delta_1 , \delta_2)\\
    \varphi_a \varphi_a = 1 + \mathcal{O} (r^{\frac{1}{2} + \delta_3})
    , \ D_\mu \varphi_a = \mathcal{O} (r^{- \frac{1}{2} + \delta_4}) , \ F_{\mu \nu} = \mathcal{O} (r^{- \frac{3}{2} + \delta_5})
    & (r \rightarrow \ 0 , \ 0 < \delta_3 , \delta_4 , \delta_5)
  \end{cases}
  .
  \label{eq:finite_condition}
\end{align}
For $D = 3$, we observe that the condition \eqref{eq:finite_condition} is not fulfilled due to the fact that $\varphi_a \varphi_a = 0$ for a single 't Hooft-Polyakov magnetic monopole \cite{tP-monopole} in the zero-size limit or a Wu-Yang magnetic monopole \cite{WY-monopole}, characterized by the following Ansatz functions:
\begin{align}
  A_0 = A_1 = \varphi_1 = \varphi_2 = 0 \ \Leftrightarrow \ \mathscr{A}_4^A (x) = 0 , \ \mathscr{A}_j^A (x) = \frac{\varepsilon_{j A k} x_k}{r^2} .
\end{align}

On the other hand, when the topological configurations depend on $t$ in the $D = 4$ case, we find that a single Alfaro-Fubini-Furlan meron \cite{AFF-meron}, specified by the following Ansatz functions:
\begin{align}
  A_0 = \frac{r}{x^2} , \ A_1 = - \frac{t}{x^2} , \ \varphi_1 = - \frac{r t}{x^2} , \ \varphi_2 = - \frac{t^2}{x^2} ,
\end{align}
results in a divergent Euclidean action, as is well known. This divergence can also be explicitly verified by performing the full integration in \eqref{eq:YM2_action}.
These topological configurations were also considered in our previous study \cite{Kondo-Fukushima}, despite their divergent Euclidean action.

For $D = 3$, it has been shown \cite{Nishino2018} that a 't Hooft-Polyakov magnetic monopole \cite{tP-monopole} with a finite size yields a finite Euclidean action, with the Ansatz function:
\begin{align}
A_0 = A_1 = \varphi_1 = 0 , \ \varphi_2 \neq 0.
\end{align}
For $D = 4$, an example of a topological configuration with a finite Euclidean action is given by the BPST instanton \cite{BPST-instanton} with a nonzero size parameter $\lambda$ \cite{Actor79}, described by the following Ansatz functions:
\begin{align}
  A_0 = F (r , t) r , \ A_1 = - F (r , t) t , \ \varphi_1 = - F (r , t) r t , \ 1 + \varphi_2 = F (r , t) r^2 , \ F(r , t) := \frac{2}{r^2 + t^2 + \lambda^2} .
  \label{eq:Instanton_Witten}
\end{align}
Thus, we can select a topological configuration that preserves the RGS while yielding a finite Euclidean action, thereby contributing meaningfully to the path integral.

\section{Conclusion and discussion}
In this talk, we tried to extend the KO color confinement criterion by including the topological configurations, formulating it as a condition for the restoration of the RGS in a single gauge field sector. Furthermore, we have examined the existence of RGS and how the restoration of RGS can be understood in terms of topological configurations that contribute to the path integral by introducing the Witten Ansatz.

For a throughout understanding of confinement, it is essential to take into account interactions among various fields, including ghost fields and gauge fields. In this context, we have provided a physical picture in which symmetry restoration occurs due to topological configurations contributing to the path integral. Based on this perspective, it is important to further investigate the condensation of topological defects in arbitrary dimensions.

Since the Witten Ansatz restricts the gauge group $SU (2)$ to the maximal torus subgroup $U (1)$, it is interesting to consider its relation to the Maximal Abelian gauge.

Witten Ansatz causes dimensional reduction from $D = 4$ $SU (2)$ Yang-Mills theory to $D = 2$ $U (1)$ gauge-scalar theory. In $D = 2$ $U (1)$ gauge-scalar theory, it is known that the RGS is restored and confinement criterion of Wilson is satisfied by the effect of instantons. It would be interesting to explore this direction further.

\section*{Acknowledgment}
This work was supported by Grant-in-Aid for Scientific Research, JSPS KAKENHI Grant Number (C) No.23K03406.
N.F. was supported by JST, the establishment
of university fellowships towards the creation of science
technology innovation, Grant Number JPMJFS2107 and Grant Number JPMJSP2109.

\end{document}